\documentclass{ws-ijmpa}

\makeatletter

\def\lsim{\compoundrel<\over\sim}
\def\compoundrel#1\over#2{\mathpalette\compoundreL{{#1}\over{#2}}}
\def\compoundreL#1#2{\compoundREL#1#2}
\def\compoundREL#1#2\over#3{\mathrel
  {\vcenter{\hbox{$\m@th\buildrel{#1#2}\over{#1#3}$}}}}
\makeatother

\begin{document}

\title{{Can The Majorana neutrino CP-violating phases be restricted? }}

\author{KOICHI MATSUDA}
\address{Department of Physics, Osaka University, \\
Toyonaka, Osaka, 560-0043, Japan\\
matsuda@het.phys.sci.osaka-u.ac.jp
}

\author{TAKESHI FUKUYAMA}
\address{Department of Physics, Ritsumeikan University, \\
Kusatsu, Shiga, 525-8577 Japan\\
fukuyama@se.ritsumei.ac.jp
}

\author{HIROYUKI NISHIURA}
\address{Department of General Education, \\
Junior College of Osaka Institute of Technology, \\
Asahi-ku, Osaka,535-8585 Japan\\
nishiura@jc.oit.ac.jp
}

\maketitle

\pub{Received (July 14, 2003)}{
}

\begin{abstract}
We reanalyze the constraints in neutrino masses and MNS 
lepton mixing parameters using the new data from the terrestrial 
(KamLAND) and astrophysical (WMAP) observations together 
with the HEIDELBERG-MOSCOW double beta decay experiment.
It leads us to the almost degenerate or inverse hierarchy 
neutrino mass scenario. 
We discuss the possibility of getting the bound for 
the Majorana $CP$ violating phase. 
\end{abstract}

\ \\
Recently the two important experimental results on neutrino physics have been 
successively released. One comes from the KamLAND \cite{eguchi} and the other does 
from the WMAP \cite{WMAP}. 
In this letter, by using these values together with Heiderberg-Moscow result 
\cite{Klapdor}, we constrain one of the two Majorana phases in the framework of our treatment \cite{matsuda}. 
The other Majorana phase cannot be restricted because of the smallness of $U_{e3}$.
We use the following experimental values.\\
(1) Heiderberg-Moscow result on the averaged neutrino mass \cite{Klapdor}
\begin{eqnarray}
\langle m_\nu \rangle
	&=& 0.39 \mbox{ [eV]} \mbox{  (best fit)} \nonumber \\
	&=& 0.11 \ - \ 0.56 \mbox{ [eV]} \mbox{  (95\%CL).}
\label{H-M}
\end{eqnarray}
(2) WMAP result on the neutrino masses \cite{WMAP}
\begin{eqnarray}
\sum_{i=1}^{3} m_i &<& 0.70 \mbox{ [eV]} \mbox{  (95\%CL).}
\label{WMAP}
\end{eqnarray}
(3) Solar neutrino \& KamLAND (l-LMA solution) \cite{Smirnov}
\begin{eqnarray}
\sin^2 2\theta_{12}
	&=& 0.82 \mbox{  (best fit),} \nonumber \\
	&=& 0.70 \ - \ 0.96 \mbox{  (95\%CL).}
\label{solar}
\end{eqnarray}
(4) CHOOZ \cite{chooz}
\begin{eqnarray}
\sin^2 \theta_{13}
	&<& 0.03 \mbox{  (90\%CL).}
\label{chooz}
\end{eqnarray}
The differences of the squared masses \(\Delta m_{ij}^2 \equiv |m_j^2 - m_i^2|\) measured
by neutrino oscillation experiments are not sensitive to our phase analysis.
Therefore we only use these best fit values \cite{Smirnov}\hspace{0.5mm}\cite{shiozawa}.
\begin{eqnarray}
\Delta m_{12}^2 &=& 7.32 \times 10^{-5} \mbox{ [eV]}^2 \quad \mbox{(l-LMA)}, 
\nonumber \\ 
\mbox{and} \quad \Delta m_{23}^2 &=& 2.5 \times 10^{-3} \mbox{ [eV]}^2 \quad 
(\mbox{Atmospheric } \nu \mbox{ exp.})
\end{eqnarray}
Moreover, we estimate very roughly the errors of 
\(\langle m_{\nu} \rangle^2\), \(\sum_{i=1}^3 m_i\), \(\sin^2 2\theta_{12}\)
and suppose the experimental data are distributed as a
normal (Gaussian) distribution around the best fit.
(\(1 \sigma = 68.3\) \% CL, \(1.45 \sigma = 85.0\) \% CL, 
\(1.65 \sigma = 90.0\) \% CL, \(1.96 \sigma = 95.0\) \% CL)
Namely, we use the following values.
\begin{eqnarray}
\langle m_\nu \rangle^2
	& = & 0.39^2 \pm (0.39^2-0.11^2) \times \frac{1.65}{1.96} 
          =   0.15 \pm 0.12 \mbox{  (90\%CL)}, \\
\sum_{i=1}^{3} m_i 
	& < & 0.00 - (0.00- 0.70) \times \frac{1.65}{1.96} = 0.59 \mbox{  (90\%CL)}, \\
\sum_{i=1}^{3} m_i 
	& < & 0.00 - (0.00- 0.70) \times \frac{1.45}{1.96} = 0.52 \mbox{  (85\%CL)}.\\
\sin^2 2\theta_{12}
	&=& 0.82 \pm (0.82- 0.70) \times \frac{1.65}{1.96} = 0.82 \pm 0.11 \mbox{  (90\%CL)}.
\end{eqnarray}
The assumptions of normal distribution in 
\(\langle m_\nu \rangle^2\), \(\sum_{i=1}^{3} m_i\)
and \(\sin^2 2\theta_{12}\) are considered to be not so bad 
from Table 2 in the paper of Klapdor-Kleingrothaus et al. \cite{Klapdor},
 Fig.1 of Hannestad \cite{Hannestad} and Fig.4 of Holand-Smirnov \cite{Smirnov}, 

Maki-Nakagawa-Sakata (MNS) mixing matrix 
$U$ takes the following form in the standard representation:
\begin{equation}
U=
\left(
\begin{array}{ccc}
c_1c_3&s_1c_3e^{i\beta}&s_3e^{i(\rho-\phi )}\\
(-s_1c_2-c_1s_2s_3e^{i\phi})e^{-i\beta}&
c_1c_2-s_1s_2s_3e^{i\phi}&s_2c_3e^{i(\rho-\beta )}\\
(s_1s_2-c_1c_2s_3e^{i\phi})e^{-i\rho}&
(-c_1s_2-s_1c_2s_3e^{i\phi})e^{-i(\rho-\beta )}&c_2c_3\\
\end{array}
\right).\label{CKM}
\end{equation}
Here $c_j=\cos\theta_j$, $s_j=\sin\theta_j$ 
($\theta_1=\theta_{12},~\theta_2=\theta_{23},~\theta_3=\theta_{31}$) \cite{bilenky}. 
Note that, for Majorana particles, there appear three 
$CP$ violating phases, the Dirac phase $\phi$ and the Majorana phases $\beta$ , $\rho$.
Irrespectively of the $CP$ violating phases we have the inequality \cite{matsuda} 
on the averaged mass,
\begin{eqnarray}
\langle m_{\nu} \rangle &\equiv& |\sum _{j=1}^{3}U_{ej}^2m_j|\\
&<& |U_{e1}|^2m_1+|U_{e2}|^2\sqrt{m_1^2+\Delta m_{12}^2}
+|U_{e3}|^2 \sqrt{m_1^2+\Delta m_{12}^2+\Delta m_{23}^2}.
\label{hierarchy}
\end{eqnarray}
Here we have used the constraint from the oscillation experiments of CHOOZ\cite{chooz} 
and SuperKamiokande\cite{skamioka}. 
It is apparent from Eqs.(\ref{H-M}), (\ref{WMAP}), and (\ref{solar}) that the normal 
hierarchy, \(m_1\lsim m_2\ll m_3\), is forbidden.  
We know that the inverse hierarchy is disfavored by the observation of 
Supernova 1987A \cite{minakata} and by the realistic GUT model \cite{fukuyama}. 
However we have no way of distinguishing between the almost degenerate 
and inverse hierarchy neutrino mass scenarios based on Eq.(\ref{H-M}) at this stage, 
because \(|U_{e1}|^2m_1\)\(+\)\(|U_{e2}|^2\sqrt{m_1^2+\Delta m_{12}^2}\)
\(\gg\)\(|U_{e3}|^2 \sqrt{m_1^2+\Delta m_{12}^2+\Delta m_{23}^2}\).
Keeping these in mind, we adopt that the neutrino masses are almost degenerate and 
\begin{eqnarray}
\langle m_{\nu} \rangle& \simeq& m||U_{e1}|^2+|U_{e2}|^2e^{2i\beta}|,
\label{betabetamass2}
\end{eqnarray}
with \(m\equiv m_1 \simeq m_2\).
Since Eq.(\ref{chooz}), \(\sin^2{2\theta_{12}}\) becomes 
\(4|U_{e2}|^2(1-|U_{e2}|^2)\) and Eq.(\ref{betabetamass2}) is rewritten as
\begin{equation}
\sin^2\beta=\frac{1}{\sin^2{2\theta_{12}}}
\left(1-\frac{\langle m_{\nu} \rangle^2}{m^2}\right) .\label{betaphase}
\end{equation}
Eq.(\ref{betaphase}) gives
\begin{equation}
\sin^2\beta \le \frac{1}{(\sin^2{2\theta_{12}})}
\left(1-\frac{\langle m_{\nu} \rangle^2}{m_{\mbox{\tiny max}}^2}\right) .\label{eq05}
\label{inequality1}
\end{equation}
Here we have denoted the experimental upper limits of \(m\) obtained from Eq.(\ref{WMAP}) 
as \(m_{\mbox{\tiny max}}\).
Let us superimpose the constraints of the other experimental bounds of Eqs. (\ref{H-M}) 
and (\ref{solar}) on this inequality in Fig.1.
When 1-dimensional restriction is translated into 2-dimensional one, 
the following region approximately coincide with \(85\%\) C.L.  
\begin{equation}
\chi^2(\sin^22\theta_{12}, \langle m_{\nu} \rangle^2) 
\equiv  \left(\frac{\sin^2 2\theta_{12}-0.82}{0.70-0.82}\right)^2
     +  \left(\frac{\langle m_{\nu} \rangle^2 - 0.39^2}{0.11^2-0.39^2}\right)^2
< 1,
\end{equation}
because we assume these values are distributed as a normal (Gaussian) distribution.
In another respect, Eq.(\ref{betaphase}) gives the upper limit of \(\sin^2\beta\) as
\begin{equation}
\sin^2\beta \le \frac{1}{(\sin^2{2\theta_{12}})\mbox{\tiny min}}
\left(1-\frac{\langle m_{\nu} \rangle\mbox{\tiny min}^2}{m^2}\right).
\end{equation}
in the confined region \(\chi((\sin^22\theta_{12})_{\mbox{\tiny min}}, 
\langle m_{\nu} \rangle_{\mbox{\tiny min}}^2)\) \(<\) 1.
Then we obtain the allowed region in the $\sin^2\beta-m$ plane in Fig.2. 
By combining this with the WMAP experiments, 
we have the meaningful constraint on the Majorana phase $\beta$ with \(\lsim 85\%\) C.L. 
for LMA-MSW solution.
Namely, we have
\begin{equation}
\sin^2 \beta \lsim 0.71
\end{equation}
at 85 \%C.L. 
And we obtain the lower limits of neutrino mass.

We must consider the factor of uncertainty in the nuclear matrix elements as well.
This uncertainty enlarges the range of \(\langle m_{\nu} \rangle\) to, 
\cite{Klapdor}
\begin{eqnarray}
\langle m_{\nu} \rangle &=& (0.05-0.84) \mbox{  eV,  (95\% C.L.)}\\
\chi^2(\sin^22\theta_{12}, \langle m_{\nu} \rangle^2) 
&\equiv&  \left(\frac{\sin^2 2\theta_{12}-0.82}{0.70-0.82}\right)^2
       +  \left(\frac{\langle m_{\nu} \rangle^2 - (0.05^2+ 0.84^2)/2}
                     {0.05^2-(0.05^2+ 0.84^2)/2}\right)^2 \nonumber \\
&<& 1.  \mbox{      (85\% C.L.)}
\end{eqnarray}
In this case, \(\sin^2 \beta\) is not restricted as shown by Fig.1 and Fig.2.
Therefore, the reliability of the HEIDELBERG-MOSCOW $(\beta\beta)_{0\nu}$ experimental 
results must be checked more precisely 
by other near future $(\beta\beta)_{0\nu}$ experiments,
 which may enable us to understand one (\(\beta\)) of two 
Majorana phases more definitely.
However, it will be difficult to measure another phase (\(\rho\)).
Finally, we must note the near future 3H beta decay experiments, 
KATRIN\cite{katrin}.
After three years of measuring time, this upper limit will be improved to
\begin{equation}
m \lsim 0.35 \mbox{[eV]  (90\%CL).}
\end{equation}
It will be very useful to get more detailed 
information about the Majorana phases and to check the mutual consistencies 
among many parameters \cite{matsuda2}.

We are grateful to express our sincere thanks to O. Yasuda for the useful comments 
on error analysis. 
This work of K.M. was supported by the JSPS Research Fellowships 
for Young Scientists, No. 3700.

\begin{figure}[htbp]
\begin{center}
\includegraphics[width=10cm]{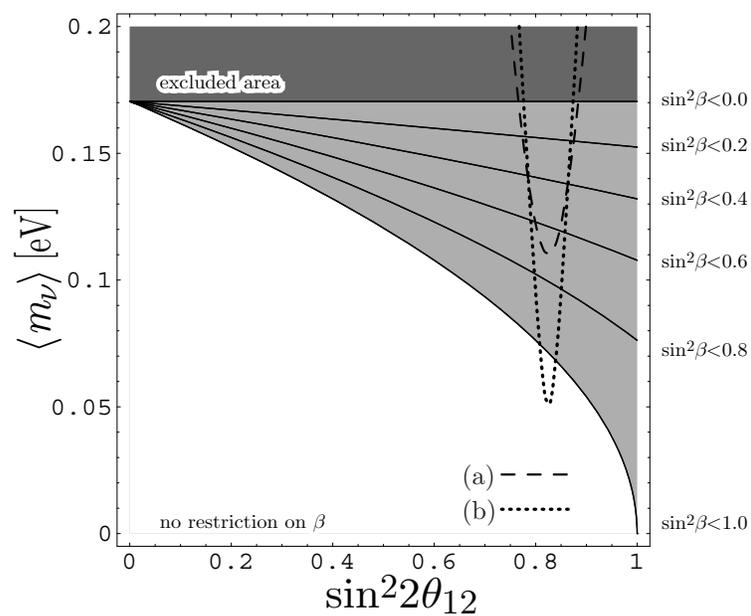}
\end{center}
\caption{The possible upper bounds for \(\sin^2\beta\) 
on the \(\sin^2 2\theta_{12}- \langle m_{\nu} \rangle\) plane.
Lines (a) and (b) show the cases where the uncertainty of the nuclear matrix elements 
is neglected and considered, respectively.
These contour lines of \(\chi^2\) indicate \(85\%\) C.L. 
from Eqs.(\ref{H-M}) and (\ref{WMAP}).}
\label{fig1}
\end{figure}

\begin{figure}[htbp]
\begin{center}
\includegraphics[width=10cm]{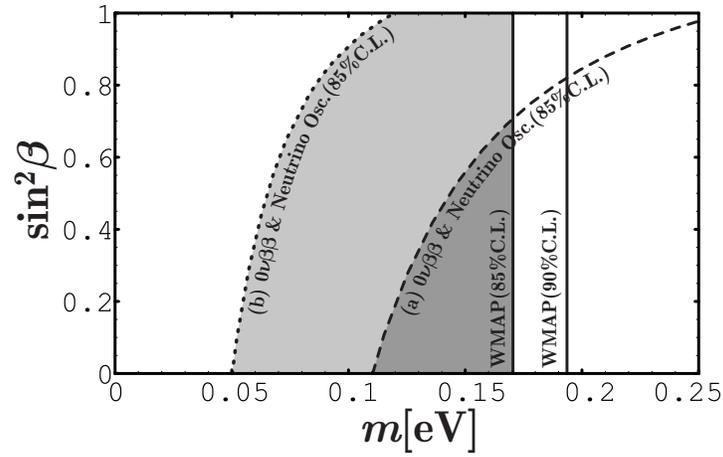}
\end{center}
\caption{
The allowed region in the \(m-\sin^2 \beta\) plane. 
(a) and (b) regions show the cases where the uncertainty of the nuclear matrix 
elements is neglected and considered, respectively.
Thus \(\sin^2\beta\) has the upper limit with 85\(\%\)C.L. 
if we do not consider the uncertainty.
}
\label{fig2}
\end{figure}

\end{document}